\documentclass[aps,prd,twocolumn,floatfix,preprintnumbers,showpacs]{revtex4}
\usepackage{amssymb,amsmath,bm,graphicx,longtable}
\usepackage{mathrsfs,enumerate}

\begin{document}
\preprint{CHIBA-EP-194/KEK Preprint 2012-10}

\title{Magnetic monopole loops generated from calorons with nontrivial holonomy}

\author{Nobuyuki Fukui$^{1}$}
\email[]{n.fukui@graduate.chiba-u.jp}

\author{Kei-Ichi Kondo$^{1}$}
\email[]{kondok@faculty.chiba-u.jp}

\author{Akihiro Shibata$^{2}$}
\email[]{akihiro.shibata@kek.jp}

\author{Toru Shinohara$^{1}$}
\email[]{sinohara@graduate.chiba-u.jp}

\affiliation{
$^1$Department of Physics, Graduate School of Science, Chiba University,
Chiba 263-8522, Japan
\\
$^2$Computing Research Center,
High Energy Accelerator Research Organization (KEK) 
and Graduate Univ. for Advanced Studies (Sokendai),
Tsukuba  305-0801, Japan
}

\date{\today}

\begin{abstract}
We study whether or not magnetic monopoles are generated from calorons defined in the space $\mathbb{R}^3\times S^1$ with the period $\beta$. 
We give numerical evidence that one-caloron solution with nontrivial holonomy generates two loops of magnetic monopole and each loop passing through one of the two poles of the caloron winds along the time direction for small $\beta$, while two loops approach each other to fuse into an unwinding loop for large $\beta$, suggesting  the existence of a critical value of $\beta$  separating two different phases.   
This work is  a first step to explain quark confinement/deconfinement at finite temperature from the viewpoint of dual superconductor picture in our framework. 

\end{abstract}

\pacs{12.38.Aw, 21.65.Qr}

\maketitle

\section{introduction}

The dual superconductor picture \cite{dualsuper} was proposed as a promising mechanism for explaining quark confinement due to strong interactions.
In this picture, quark confinement is realized by squeezing   color electric fields connecting quark and antiquark due to the dual Meissner effect which  originates from condensation of magnetic monopoles.
For this picture to be true, there must exist a Yang-Mills field configuration from which a magnetic monopole originated  draws a trajectory of a closed loop in four-dimensional spacetime  (as guaranteed from the magnetic current conservation) \cite{Kondo08b}, while a magnetic monopole is a point-like object in three-dimensional spacetime. 
A candidate for such a Yang-Mills field configuration will be the classical solution of the Yang-Mills field equation, which is expected to give a dominant contribution to the path-integral of the Yang-Mills theory.  Moreover, we look for the Yang-Mills field having a nontrivial topological invariant with a desire that it could be related to the magnetic charge in the dual description of the Yang-Mills theory.

From this viewpoint, we have studied \cite{KFSS08,SKKISF09,FKSS10} topological classical solutions of the Yang-Mills field equation; merons \cite{AFF76,AFF77,Actor79,CDG78} and instantons, and we have shown that closed loops of magnetic monopoles are indeed generated from topological classical  solutions of the Yang-Mills theory in the gauge-invariant way. 
In \cite{KFSS08}, it has been analytically shown that the 2-meron (dimeron)  as a non-self-dual solution generates closed loops of magnetic monopole, which go through two poles of the 2-meron, see also \cite{SKKISF09} for the corresponding numerical result. 
This result confirms the preceding results of numerical studies for merons performed in a specific gauge fixing \cite{MN02}.

According to preceding studies \cite{BOT97,CG95,BHVW01}, on the other hand, it was believed that instantons as self-dual solutions of the Yang-Mills field equation don't contribute to quark confinement.  In fact, one-instanton solution and multi-instanton solution of 't Hooft type failed to detect the loop of magnetic monopole. 
In \cite{FKSS10}, however, we have discovered in a numerical way that a loop of magnetic monopole is generated from the two-instanton solution of the  Jackiw-Nohl-Rebbi (JNR) type.  Moreover, we have clarified  in \cite{FKSS12} why the two-instanton of 't Hooft type does not generate the magnetic monopole loop, using a fact that the two-instanton of 't Hooft type is obtained as a special limit of JNR solution.

In this paper, we proceed to study the magnetic monopole content in caloron solutions, i.e., periodic instantons, with trivial holonomy \cite{HS78} and nontrivial holonomy \cite{KvB98,LL98}.
We study  in a numerical way whether or not the caloron can be a source for loops of magnetic monopoles defined in the space $\mathbb{R}^3\times S^1$ with the period $\beta$. 
Especially, we focus on the $\beta$ dependence on the behavior of the generated loops. 
The physical motivation of this study is to understand the confinement/deconfinement phase transition in the Yang-Mills theory at finite temperature from the viewpoint of magnetic monopoles.  This work is a first step toward this direction.

This paper is organized as follows. 
In section II, we review calorons with trivial holonomy and    nontrivial holonomy. 
In section III, we set up our framework to study the gauge-invariant magnetic monopole derived from a given Yang-Mills field configuration, and explain how to perform numerical calculations on a lattice used for regularization. 
In section IV, we give a main result of our numerical calculations for detecting loops of magnetic monopole.
In the final section, we summarize the result. 

\section{caloron}

A caloron is a solution of the self-dual equation for 
the Yang-Mills field ${\bf A}$ on $\mathbb{R}^3\times S^1$.
The self-dual equation is given by
\begin{equation}
 {}^\ast{\bf F}_{\mu\nu}(x)={\bf F}_{\mu\nu}(x) ,
\end{equation}
where ${\bf F}_{\mu\nu}$ is the field strength of ${\bf A}_\mu$:
\begin{gather}
 {\bf F}_{\mu\nu}(x) 
= \partial_\mu{\bf A}_\nu(x) - \partial_\nu{\bf A}_\mu(x)
             -ig\left[{\bf A}_\mu(x) , {\bf A}_\nu(x) \right] ,
\end{gather}
and ${}^\ast{\bf F}_{\mu\nu}$ is the Hodge dual of ${\bf F}_{\mu\nu}$:
\begin{equation}
 {}^\ast{\bf F}_{\mu\nu}(x)
 =\frac{1}{2}\epsilon_{\mu\nu\rho\sigma}{\bf F}_{\rho\sigma}(x) .
\end{equation} 
Here the periodic boundary condition should be understood on the gauge field ${\bf A}_\mu(x)$ defined on $\mathbb{R}^3\times S^1$:
\begin{gather}
 {\bf A}_\mu(\vec{x},t+\beta)={\bf A}_\mu(\vec{x},t),\label{periodicBC}\\
  \vec{x}\in\mathbb{R}^3,\quad t\in S^1,\quad x =\left(\vec{x},t\right) ,
\end{gather}
where the periodicity $\beta$ in Eq.\eqref{periodicBC} is the circumference of $S^1$.

A caloron is classified by the topological charge $Q$ with the charge density $D$:
\begin{align}
 Q&=\int d^3x\int_0^\beta dt\; D(\vec{x},t)\notag\\
  &:=\frac{1}{16\pi^2}\int d^3x\int_0^\beta dt\;
   \text{tr}\left[{\bf F}_{\mu\nu}(\vec{x},t)\;
                  {}^\ast\!{\bf F}_{\mu\nu}(\vec{x},t)\right] ,
 \label{instanton charge in TYM}
\end{align}
just like an instanton.
In addition, it is specified by a holonomy:
\begin{equation}
 H=\lim_{|\vec{x}|\rightarrow\infty}
   P\exp\left\{ig\int_0^\beta dt\;{\bf A}_4(\vec{x},t)\right\}
   \label{def:holonomy} ,
\end{equation}
where $P$ represents a path-ordered product along $S^1$.
The existence of the periodicity $\beta$ and the holonomy are new aspects of calorons, compared with instantons.

\subsection{HS caloron}

The  one-caloron  having a unit topological charge $Q=1$ and a trivial holonomy  $H=\bm{1}$ 
for the gauge group $SU(2)$ 
was discovered by B. J. Harrington and H. K. Shepard \cite{HS78}.
This solution is called the HS caloron.  
The HS caloron is given by
\begin{align}
 & g{\bf A}_\mu(\vec{x},t)
  =-\eta_{\mu\nu}^{A(\mp)}T^A\partial_\nu\log\phi(\vec{x},t), \notag\\
& \phi(\vec{x},t)
 =1+\frac{\lambda\rho^2}{2\left|\vec{x}-\vec{x}_0\right|}
     \frac{\sinh\left(\lambda\left|\vec{x}-\vec{x}_0\right|\right)}
          {\cosh\left(\lambda\left|\vec{x}-\vec{x}_0\right|\right)
          -\cos\left(\lambda\left(t-t_0\right)\right)}\label{one-caloron},
\end{align}
where $T_A=\sigma_A/2$ ($\sigma_A$: Pauli matrices)
and $\eta_{\mu\nu}^{A(\pm)}$ is the symbol defined by
\begin{equation}
 \eta_{\mu\nu}^{A(\pm)}=\epsilon_{A\mu\nu4}
                        \pm
                        \delta_{A\mu}\delta_{\nu4}
                        \mp
                        \delta_{A\nu}\delta_{\mu4} .
\end{equation}
Here $x_0^\mu=\left(\vec{x}_0,t_0\right)$ is the center parameter 
and $\rho$ is the size parameter, and $\lambda$ is the parameter associated with $\beta$ by $\lambda=2\pi/\beta$.

\subsection{KvBLL caloron}

The one-caloron having a unit topological charge $Q=1$ and a nontrivial holonomy $H \not= \bm{1}$ for the gauge group $SU(2)$ was found out
by C. Kraan and van Baal \cite{KvB98} and independently by Lee and Lu
\cite{LL98}.
This solution is called the KvBLL caloron.
The KvBLL caloron is given by
\begin{align}
g{\bf A}_\mu
 &=-\left(\eta_{\mu\nu}^{3(+)}\partial_\nu\log\Phi+v\delta_{\mu4}\right)T_3
 \notag\\
 &
  -\Phi\text{Re}\!\left[\left(\eta_{\mu\nu}^{1(+)}
                             -i\eta_{\mu\nu}^{2(+)}\right)
                       \left(T_1+iT_2\right)
                       \left(\partial_\nu+iv\delta_{\nu4}\right)\zeta
                 \right] ,
\label{periodic KvBLL}
\end{align}
where $\text{Re}M=(M+M^\dagger)/2$ for the matrix $M$.
The used functions are
\begin{align}
 \Phi&=\frac{\psi}{\hat\psi} ,
\\
 \hat\psi&=-\cos(\mu(t-t_0))
          +\cosh(w|\vec{r}|)\cosh(v|\vec{s}|)\notag\\
         &\hspace{3cm}
          +\frac{\vec{r}\cdot\vec{s}}{|\vec{r}||\vec{s}|}
           \sinh(w|\vec{r}|)\sinh(v|\vec{s}|)\label{hatpsi}, 
\\
 \psi&=\hat\psi
      +\frac{\mu^2\rho^4}{4|\vec{r}||\vec{s}|}\sinh(w|\vec{r}|)
                                              \sinh(v|\vec{s}|)\notag 
\\
     &\hspace{2cm}
      +\frac{\mu\rho^2}{2s}\cosh(w|\vec{r}|)\sinh(v|\vec{s}|)\notag 
\\
     &\hspace{2cm}
      +\frac{\mu\rho^2}{2|\vec{r}|}\sinh(w|\vec{r}|)\cosh(v|\vec{s}|) ,
\\
 \zeta&=\frac{\mu\rho^2}{2\psi}
        \left(e^{-i\mu(t-t_0)}\frac{\sinh(v|\vec{s}|)}{|\vec{s}|}
       +\frac{\sinh(w|\vec{r}|)}{|\vec{r}|}\right)\label{zeta} ,
\\
 \mu&=\frac{2\pi}{\beta},\quad v=\frac{\theta}{\beta},\quad
 w=\frac{2\pi-\theta}{\beta}=\mu-v ,
\\
 \vec{r}&=\vec{x}-\vec{x}_0+\frac{\rho^2}{2\beta}\vec{\theta},\quad
 \vec{s}=\vec{x}-\vec{x}_0-\frac{\rho^2}{2\beta}\frac{w}{v}\vec{\theta},
 \\
 \vec{\theta}&=\left(0,0,\theta\right)\label{pole of KvBLL} ,
\end{align}
where $x_0^\mu=\left(\vec{x}_0,t_0\right)$ and $\rho$ are respectively the center and the size parameters, and $\theta$ is the parameter related to
the holonomy.
The KvBLL caloron is characterized by the parameters $x_0, \rho, \beta, \theta$.

As $x_\mu$ goes to spatial infinity, the KvBLL caloron Eq.\eqref{periodic KvBLL} behaves
\begin{equation}
 g{\bf A}_i(\vec{x},t) \rightarrow0, \quad 
 g{\bf A}_4(\vec{x},t) \rightarrow -v T_3
 \quad(\vec{x}\rightarrow\infty).
\end{equation}
Therefore, a holonomy of the KvBLL caloron reads
\begin{equation}
 H
 =e^{-i\beta v T_3}
 =e^{-i\theta T_3}\label{holonomy} ,
\end{equation}
which reduces to the HS caloron at $\theta=0$.

The KvBLL caloron has two poles at
$p_1=(\vec{x}_0-\left(\rho^2/2\beta\right)\vec{\theta},\  t_0)$
and
$p_2=(\vec{x}_0+\left(\rho^2/2\beta \right) \left(w/v\right)\vec{\theta},\ t_0)$.
The distance between two poles is given by
\begin{align}
 \left|p_1-p_2\right|&=
 \left|\vec{x}_0-\frac{\rho^2}{2\beta}\vec{\theta}
      -\left(\vec{x}_0+\frac{\rho^2}{2\beta}\frac{w}{v}\vec{\theta}\right)
 \right|
 =\frac{\pi\rho^2}{\beta}.
\end{align}
The KvBLL caloron behaves as a dyon in the vicinity of a respective pole.
The dyon is a BPS monopole specified by
chromo-electric and chromo-magnetic charges
 in $SU(2)$ Yang-Mills theory on $\mathbb{R}^3$ \cite{Diakonov_NPB195_5}.

\section{Setting up numerical calculation}

We proceed to extract magnetic monopoles from a caloron  using 
 a method which was originally formulated in the continuum by Cho \cite{Cho80} and  Duan and Ge \cite{DG79} independently, readdressed later by Faddeev and Niemi \cite{FN99} and Shabanov \cite{Shabanov99},  and further developed by our group \cite{KMS06,KMS05,Kondo06,KSM08}.
This method enables one to extract  magnetic monopoles
from the original Yang-Mills field without breaking the gauge symmetry.
For $SU(2)$, this method \cite{KMS06,KMS05,Kondo06} is a gauge-invariant extension of the Abelian projection
invented by 't Hooft \cite{tHooft81}.
For $SU(3)$, this method does not necessarily reduce to the Abelian projection and there appear non-Abelian magnetic monopoles, see \cite{KSM08} for the details. 

In this method for $SU(2)$, we introduce a color field ${\bf n}(x)$ with unit length into
the original Yang-Mills theory:
\begin{gather}
 {\bf n}(x)=n^A(x)T^A,\quad\left(T^A :=\sigma_A/2\right) ,
\\
  n^A(x)n^A(x)=1 ,
\end{gather}
where $\sigma_A\ (A=1,2,3)$ are the Pauli matrices.
The following field ${\bf V}_\mu(x)$ defined from the original Yang-Mills field ${\bf A}_\mu(x)$ and the 
additional color field
${\bf n}(x)$ plays the key role in this formulation:
\begin{equation}
 {\bf V}_\mu(x) :=
c_\mu(x) {\bf n}(x)-ig^{-1}\left[\partial_\mu{\bf n}(x), {\bf n}(x)\right] ,
\label{V}
\end{equation}
where 
\begin{equation}
 c_\mu(x) :=2 \text{tr}\left({\bf n}(x){\bf A}_\mu(x)\right) .
\end{equation}
In fact, from the remarkable fact that the field strength of ${\bf V}_\mu(x)$ is parallel to ${\bf n}(x)$:
\begin{align}
 {\bf F}_{\mu\nu}[{\bf V}]
 &=\partial_\mu{\bf V}_\nu
  -\partial_\nu{\bf V}_\mu
  -ig\left[{\bf V}_\mu, {\bf V}_\nu\right]\notag\\
 &=\left\{\partial_\mu c_\nu
         -\partial_\nu c_\mu
         +2ig^{-1}
          \text{tr}
          \big({\bf n}
                \left[\partial_\mu{\bf n}, \partial_\nu{\bf n}\right]
          \big)
   \right\}{\bf n}\notag
\\
 &:= G_{\mu\nu}{\bf n},
\end{align}
we can define the gauge-invariant field strength 
\begin{equation}
G_{\mu\nu}(x) = 2\text{tr}({\bf n}  {\bf F}_{\mu\nu}[{\bf V}]) ,
\end{equation}
and the gauge-invariant monopole current as 
\begin{equation}
 k^\mu(x)
 :=  \partial_\nu\!\,^\ast G^{\mu\nu}(x)\label{k}
 =  \frac{1}{2}\epsilon^{\mu\nu\rho\sigma}\partial_\nu G_{\rho\sigma}(x) .
\end{equation}

In order for this method to work, the color field must be determined from the original Yang-Mills field. 
For this purpose, we impose a condition 
of minimizing the functional, which we call the reduction functional:
\begin{equation}
F_{\text{red}} 
:= \int d^3x \int_0^\beta d t\frac12 \text{tr} [\{D_\mu[{\bf A}]{\bf n}(x)\}^2   ],
\end{equation}
where
$D_\mu[{\bf A}]=\partial_\mu-ig{\bf A}_\mu$.
We can obtain ${\bf n}(x)$ 
by solving a differential equation which
we call the reduction differential equation (RDE):
\begin{equation}
 - D_\mu[{\bf A}]D_\mu[{\bf A}]{\bf n}(x)=\lambda(x){\bf n}(x) .
\label{RDE}
\end{equation}
The solutions of the RDE give the local minima of the functional $F_{\text{red}}$.
Here we impose the periodic boundary condition:
\begin{equation}
 {\bf n}(\vec{x},t+\beta)={\bf n}(\vec{x},t) ,
\end{equation}
since ${\bf n}(x)$ must be defined on $\mathbb{R}^3\times S^1$.

Once the reduction condition is solved, thus,  
we can obtain the monopole current $k^\mu$ from the original gauge field ${\bf A}_\mu$.
The gauge-invariant magnetic charge is defined by 
\begin{equation}
 q_m 
 :=  \int d^3 \sigma_\mu k^\mu(x)  ,
\end{equation}
in a Lorentz (or Euclidean rotation)  invariant way \cite{Kondo08b}.

In this paper,  we adopt a lattice version \cite{KKMSSI05,IKKMSS06,SKKMSI07,SKKMSI07b,KKSSI09,KSSMKI08,SKS09} of the above method and we  perform the procedures explained in the above in a numerical way. 
In the lattice regularization we use for numerical calculations, the link variable $U_{x,\mu}$ is related to a gauge field ${\bf A}_\mu$ in a continuum theory by
\begin{equation}
 U_{x,\mu}=P
            \exp
            \left\{ig\int_x^{x+a\hat{\mu}}dy{\bf A}_\mu(y)\right\} ,
 \label{U}
\end{equation}
where $a$ is a lattice spacing and
$\hat{\mu}$ represents the unit vector in the $\mu$ direction.
The lattice version of the reduction functional in $SU(2)$ Yang-Mills theory is given by  
\begin{equation}
 F_{\text{red}}[{\bf n},U]
 =\sum_{x,\mu}
  \left\{1-4\,\text{tr}
          \left(U_{x,\mu}{\bf n}_{x+a\hat{\mu}}U_{x,\mu}^\dagger{\bf n}_x\right)
          /\text{tr}\left({\bf 1}\right)
  \right\},
\label{reduction}
\end{equation}
where ${\bf n}_x$ is a unit color field defined on a site $x$,
\begin{equation}
 {\bf n}_x=n_x^AT^A,\quad n^A_xn^A_x=1 .
 \label{lattice-n}
\end{equation}

We introduce the Lagrange multiplier $\lambda_x$ to incorporate the constraint of unit length for the color field (\ref{lattice-n}). 
Then the stationary condition for the reduction functional is given by
\begin{equation}
  \frac{\partial }{\partial n_x^A} \left\{ F_{\text{red}}[{\bf n},U] - \frac12  \sum_{x} \lambda_x (n_x^A n_x^A -1) \right\}  =0 .
\end{equation}
When $F_{\text{red}}$ takes a local minimum for a given and fixed configurations $U_{x,\mu}$, therefore, a Lagrange multiplier $\lambda_x$  satisfies
\begin{equation}
 W_x^A=\lambda_x n_x^A
\label{RDEonLattice1},
\end{equation}
and the color field $n^A_x$ satisfies
\begin{equation}
  n^A_xn^A_x=1 ,
\end{equation}
where
\begin{align}
  &W_x^A=4\sum_{\mu=1}^4\text{tr}
         \Big(U_{x,\mu}{\bf n}_{x+a\hat{\mu}}U_{x,\mu}^\dagger T^A\notag\\
  &\hspace{2cm}
        +U_{x-a\hat{\mu},\mu}T^A
               U_{x-a\hat{\mu},\mu}^\dagger{\bf n}_{x-a\hat{\mu}}
         \Big)/\text{tr}\left({\bf 1}\right)\label{W_x^A}.
\end{align}
Eq.\eqref{RDEonLattice1} is a lattice version of the  RDE.
We are able to eliminate the Lagrange multiplier to rewrite (\ref{RDEonLattice1}) into
\begin{equation}
 n_x^A=\frac{W^A_x}{\sqrt{W^B_xW^B_x}} .
\label{RDEonLattice2}
\end{equation}
A derivation of this equation is given in our previous paper \cite{FKSS10}.
Thus the color field configurations ${\bf n}_x$  are obtained by solving  \eqref{RDEonLattice2} in a numerical way.

After obtaining the ${\bf n}_x$ configuration for  given configurations $U_{x,\mu}$ in this way, we introduce a new link variable $V_{x,\mu}$ on a lattice corresponding to the restricted field (\ref{V}) by
\begin{align}
 V_{x,\mu}=& \frac{L_{x,\mu}}
                 {\sqrt{\displaystyle{\frac{1}{2}}
                        \ \text{tr}\!
                        \left[L_{x,\mu}L_{x,\mu}^\dagger\right]}} ,
\nonumber\\
 L_{x,\mu}:=& U_{x,\mu}+{\bf n}_xU_{x,\mu}{\bf n}_{x+a\hat{\mu}} .
\end{align}
Finally, the monopole current  $k_{x,\mu}$ on a lattice
is constructed  as 
\begin{equation}
 k_{x,\mu}
 =\sum_{\nu,\rho,\sigma}\frac{\epsilon_{\mu\nu\rho\sigma}}{4\pi}
  \frac{\Theta_{x+a\hat{\nu},\rho\sigma}[{\bf n},V]
       -\Theta_{x,\rho\sigma}[{\bf n},V]}{a} ,
 \label{definition_of_k}
\end{equation}
through the angle variable of the plaquette variable:
\begin{align}
 &\Theta_{x,\mu\nu}[{\bf n},V]\notag\\
 &=a^{-2}\arg
  \Big(\text{tr}
        \left\{\left({\bf 1}+{\bf n}_x\right)
               V_{x,\mu}V_{x+a\hat{\mu},\nu}
               V_{x+a\hat{\nu},\mu}^\dagger V_{x,\nu}^\dagger
        \right\}\notag\\
 &\hspace{6.2cm}/\text{tr}\left({\bf 1}\right)
  \Big) .
\end{align}
In this definition, $k_{x,\mu}$ takes an integer value \cite{KKMSSI05,IKKMSS06}.

To obtain the ${\bf n}_x$ configuration satisfying \eqref{RDEonLattice2}, 
we recursively apply \eqref{RDEonLattice2} to ${\bf n}_x$ on each site $x$
and update it keeping ${\bf n}_x$ fixed at a spatial
 boundary $\partial V_{\mathbb{R}^3}$ of a finite lattice
\begin{equation}
 V=V_{\mathbb{R}^3}\times V_{S^1} , \quad
  V_{\mathbb{R}^3}=\left[-aL_x, aL_x\right]^3, \
 V_{S^1}=\left[0, aL_t\right] ,
 \label{def-lattice}
\end{equation}
until $F_{\text{red}}$ converges.
We should impose a periodic boundary condition at $\partial V_{S^1}$.
Since we calculate the $k_{x,\mu}$ configuration for the caloron
configuration in this paper, we need to decide a boundary condition of the ${\bf n}_x$ configuration in the caloron case.
We recall that one-caloron configuration
approaches  a pure gauge at spatial infinity $|\vec{x}| \rightarrow \infty$:
\begin{equation}
 g{\bf A}_\mu(\vec{x},t)\rightarrow ih^\dagger(\vec{x},t)\partial_\mu h(\vec{x}, t) + O(|\vec{x}|^{-2}) .
\label{BehaviorOfA}
\end{equation}
Then,  ${\bf n}(x)$ as a solution of the reduction condition is supposed to behave asymptotically as
\begin{equation}
 {\bf n}(\vec{x},t)\rightarrow h^\dagger(\vec{x},t)T_3 h(\vec{x},t) + O(|\vec{x}|^{-\alpha}) ,
\label{BehaviorOfn}
\end{equation}
for a certain value of $\alpha>0$.
Under this idea, we adopt a boundary condition:
\begin{equation}
 {\bf n}_x^\text{bound} :=
 h^\dagger(\vec{x},t)T_3 h(\vec{x},t) , \ \vec{x} \in \partial 
V_{\mathbb{R}^3}.
\label{n-boundary-c}
\end{equation}
In practice,
we start with an initial state of the $\{{\bf n}_x\}$ configuration:
${\bf n}_x^\text{init}=h^\dagger(x)T_3 h(x)$ for $x \in V$. Then, we 
repeat updating ${\bf n}_x$ on each site $x$ according to \eqref{RDEonLattice2} except for the configuration ${\bf n}_x^\text{bound}$ on the boundary $\partial V$ satisfying (\ref{n-boundary-c}).

It should be remarked that these asymptotic forms (\ref{BehaviorOfA}) and (\ref{BehaviorOfn}) satisfy the RDE asymptotically in the sense that
\begin{equation}
  D_\mu[{\bf A}]{\bf n}(x) \rightarrow 0  \quad
 (|\vec{x}| \rightarrow \infty ),
\end{equation}
together with
\begin{equation}
  \lambda(x) \rightarrow 0 \quad (|\vec{x}| \rightarrow \infty),
\end{equation}
which is necessary to obtain a finite value for the reduction functional \cite{KFSS08}:
\begin{equation}
F_{\text{red}} 
= \int d^3x\int_0^\beta dt \frac12  \lambda(x) < \infty  .
\end{equation}

\section{main result}

\begin{figure*}[htbp]
 \unitlength=0.001in
 \begin{picture}(7000,5000)(0,0)
  \put(0,2500){\includegraphics[trim=0 0 0 0, width=90mm]%
              {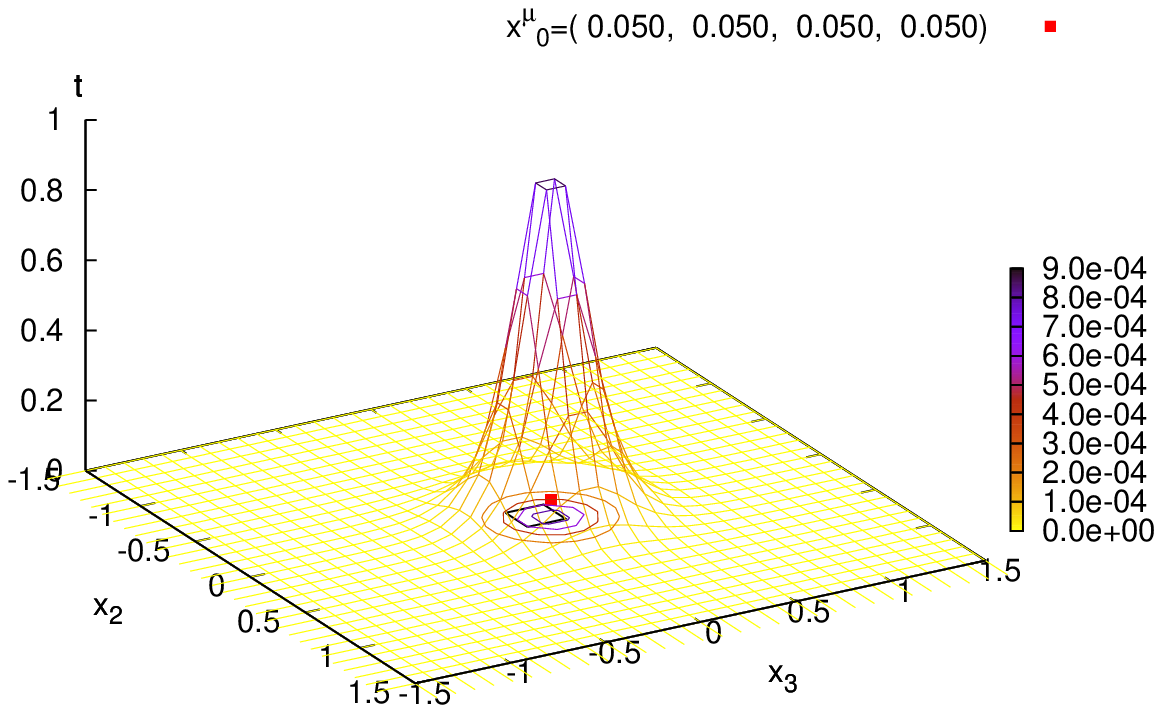}}%
  \put(1400,2500){(a)}
  \put(3500,2500){\includegraphics[trim=0 0 0 0, width=90mm]%
                 {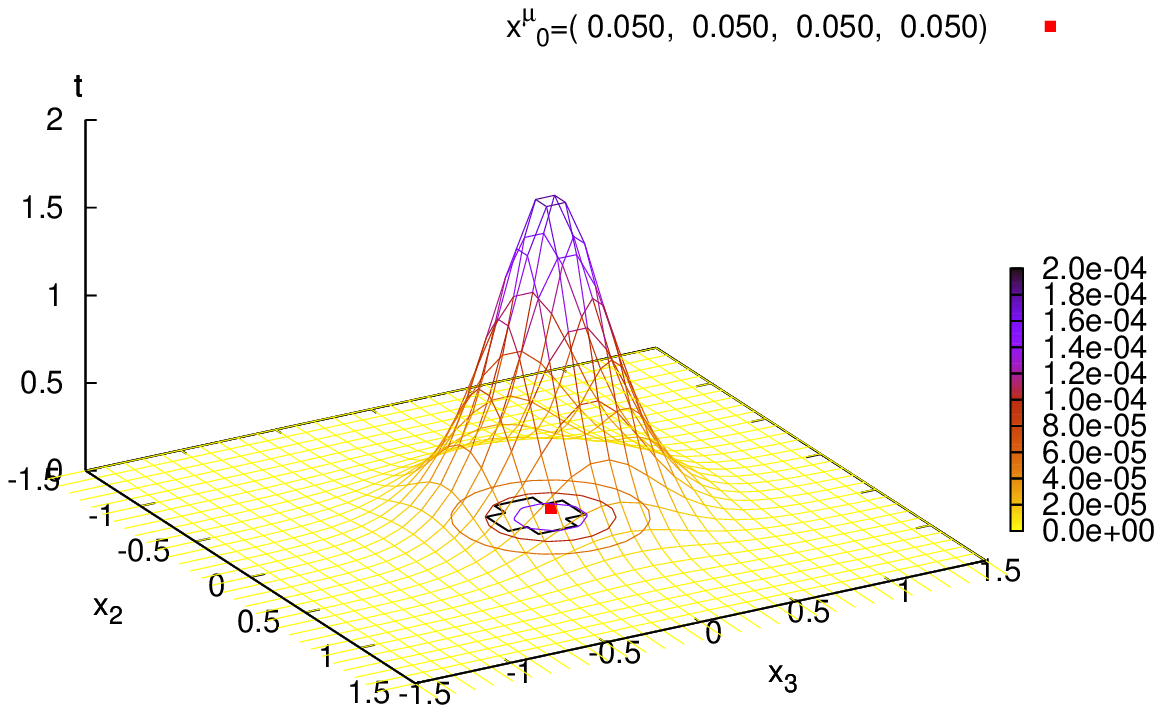}}%
  \put(4900,2500){(b)}
  \put(0,00){\includegraphics[trim=0 0 0 0, width=90mm]%
           {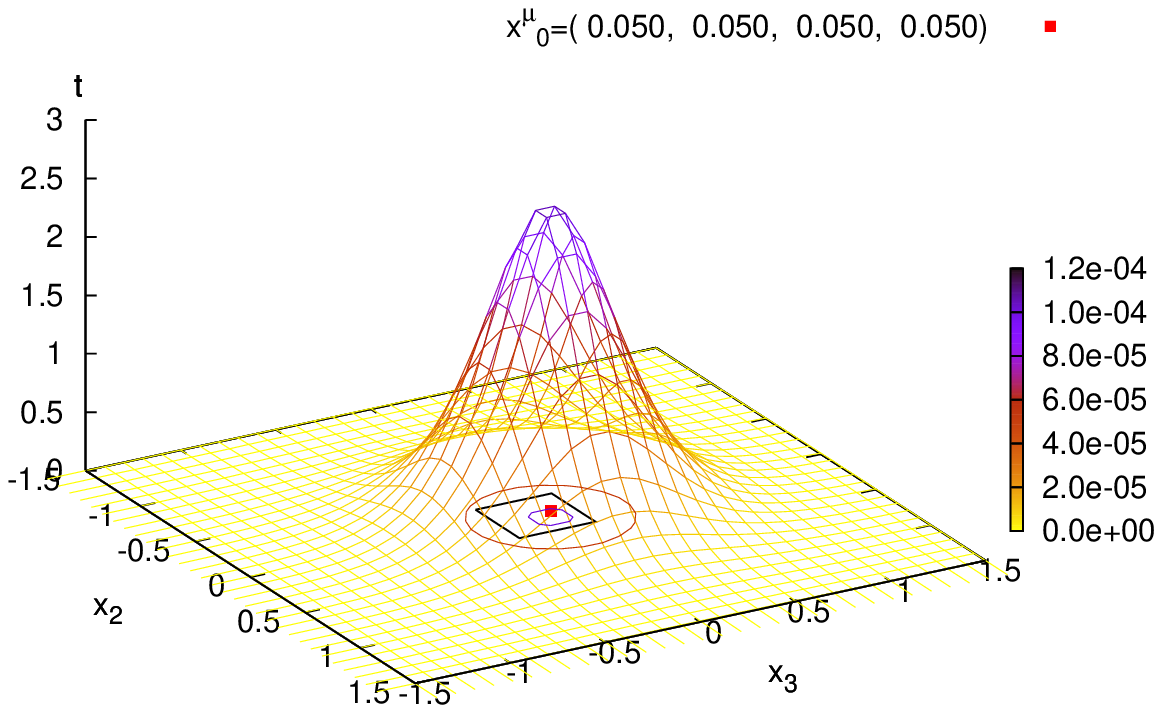}}%
  \put(1400,00){(c)}
  \put(3500,00){\includegraphics[trim=0 0 0 0, width=90mm]%
              {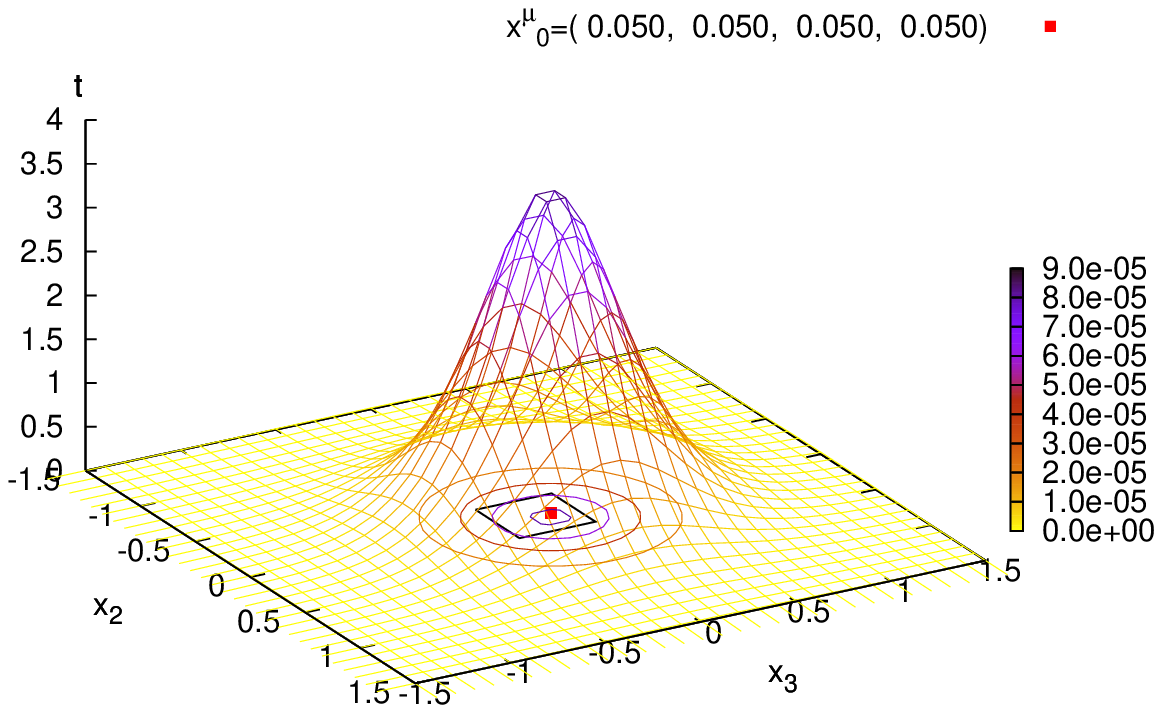}}%
  \put(4900,00){(d)}
 \end{picture}
 \caption{The charge distribution $D_x$ of the HS caloron with a fixed $\rho=10a=1.0$  and
the associated magnetic--monopole current $k_{x,\mu}$   in the three-dimensional space $x_1=0$ projected from the four-dimensional space,  which is regularized by the lattice of a finite volume $V$ with a fixed $L_x=35$ and various choices of $L_t$: (a)$L_t=10$, (b)$L_t=20$, (c)$L_t=30$,
(d)$L_t=40$.
The grid shows the caloron charge density $D_x$ on the $x_2$-$x_3$ ($x_1=t=0$) plane, together with its value being indicated in the legend on the right  side.The black (thick) curve on the $x_2$-$x_3$ plane shows the  loop of the magnetic current $k_{x,\mu}$.
The colored (thin) lines on the  $x_2$--$x_3$ plane  show a contour plot for the equi-$D_x$ lines.
Due to the  periodicity of the $t$ direction with the period $\beta=aL_t$, the top plane $t=\beta$ should be identified with the bottom one $t=0$.
Here $x_0^\mu$ denote the center parameter of the HS caloron.
}
 \label{fig:trivial1carolon}
\end{figure*}

We focus our interest on the support of $k_{x,\mu}$, namely, a set of links $\{ x,\mu \}$ on which $k_{x,\mu}$ takes non-zero values $k_{x,\mu} \not= 0$.  This locates the magnetic monopole current generated for a given Yang-Mills field configuration. 
By the definition \eqref{definition_of_k} for $k_{x,\mu}$, the number of configurations
$k_{x,\mu}$ is equal to $(2L_x)^3 \times L_t \times 4$ for the lattice specified by (\ref{def-lattice}).

We set up our numerical calculations as follows.
First, we prepare a lattice with a finite volume $V$.
In this case, the function $h(x)$ in Eq.\eqref{BehaviorOfA} is chosen as 
\begin{equation}
 h(x)=\bm{1} , 
\end{equation}
and therefore,
a boundary condition for ${\bf n}_x$ is given by
\begin{equation}
 {\bf n}_x^\text{bound}=T_3 .
 \label{n-boundary}
\end{equation}
Second, in order to define the parameters $x_0$ and $\rho$ of the caloron on the lattice, 
we fix the lattice spacing:
\begin{equation}
  a=0.1. 
\end{equation}
Then we set 
\begin{equation}
\rho=10a=1.0 , 
\end{equation}
and fix the center on 
\begin{equation}
x_0^\mu=(0,0,0,0)+\Delta ,
\end{equation}
where $\Delta$ 
is a small parameter introduced to avoid the pole singularities
at $x_0$. 
Here we have chosen:
\begin{equation}
 \Delta=(0.5a,\ 0.5a,\ 0.5a,\ 0.5a) .
\end{equation}

Third, we define the caloron charge $Q_V$ in a finite volume $V$ calculated from the $U_{x,\mu}$ configuration as
\begin{gather}
 Q_V=\sum_{x\in V-\partial V_{\mathbb{R}^3}}D_x,\nonumber\\
 D_x\equiv\frac{1}{2^4}\frac{1}{32\pi^2}
     \sum_{\mu\nu\rho\sigma=\pm1}^{\pm4}
     \hat{\epsilon}^{\mu\nu\rho\sigma}U_{x,\mu\nu}U_{x,\rho\sigma} , \nonumber\\
 U_{x,\mu\nu}=U_{x,\mu}U_{x+\hat{\mu},\nu}
              U_{x+\hat{\nu},\mu}^\dagger U_{x,\nu}^\dagger ,              
\end{gather}
where
$D_x$ is the caloron charge density  and $\hat{\epsilon}$ is related to
the usual $\epsilon$ tensor ($\text{sgn}(\mu)=\mu/|\mu|$) by
\begin{equation}
  \hat{\epsilon}^{\mu\nu\rho\sigma}
 =\text{sgn}(\mu)\text{sgn}(\nu)\text{sgn}(\rho)\text{sgn}(\sigma)
  \epsilon^{|\mu||\nu||\rho||\sigma|}  .
\end{equation}

\subsection{HS caloron}

\begin{figure*}[t]
 \unitlength=0.001in
 \begin{picture}(7000,5000)(0,0)
  \put(0,2500){\includegraphics[trim=0 0 0 0, width=90mm]%
              {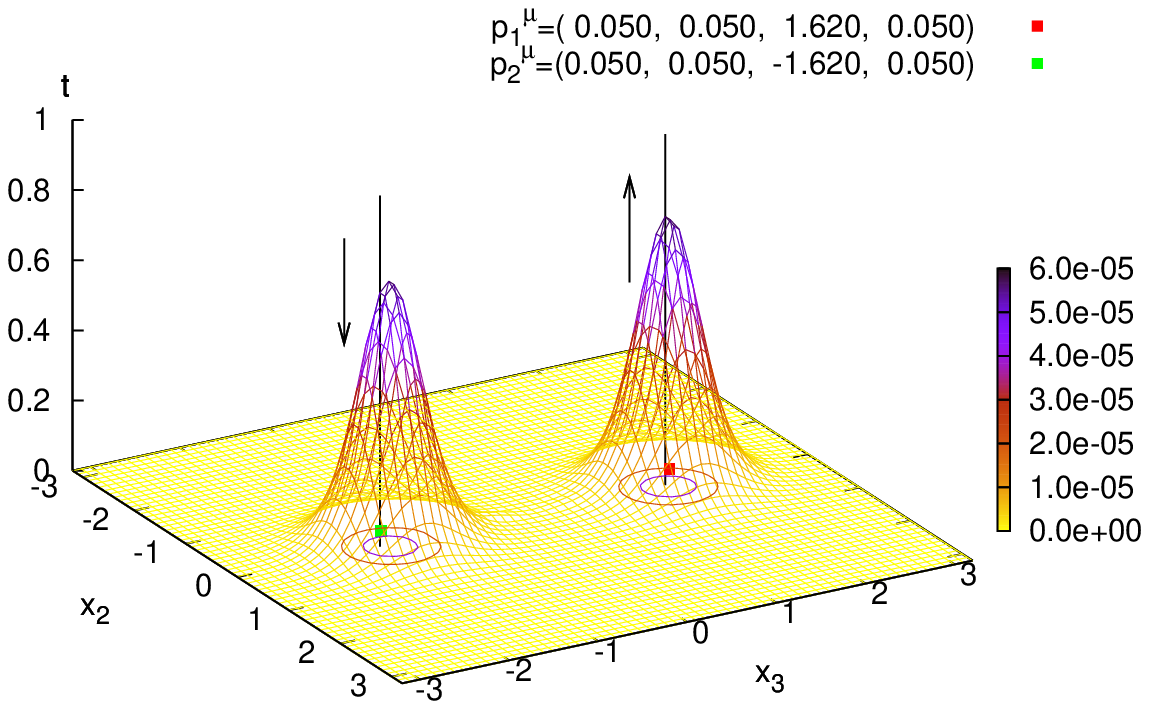}}%
  \put(1400,2500){(a)}
  \put(3500,2500){\includegraphics[trim=0 0 0 0, width=90mm]%
                 {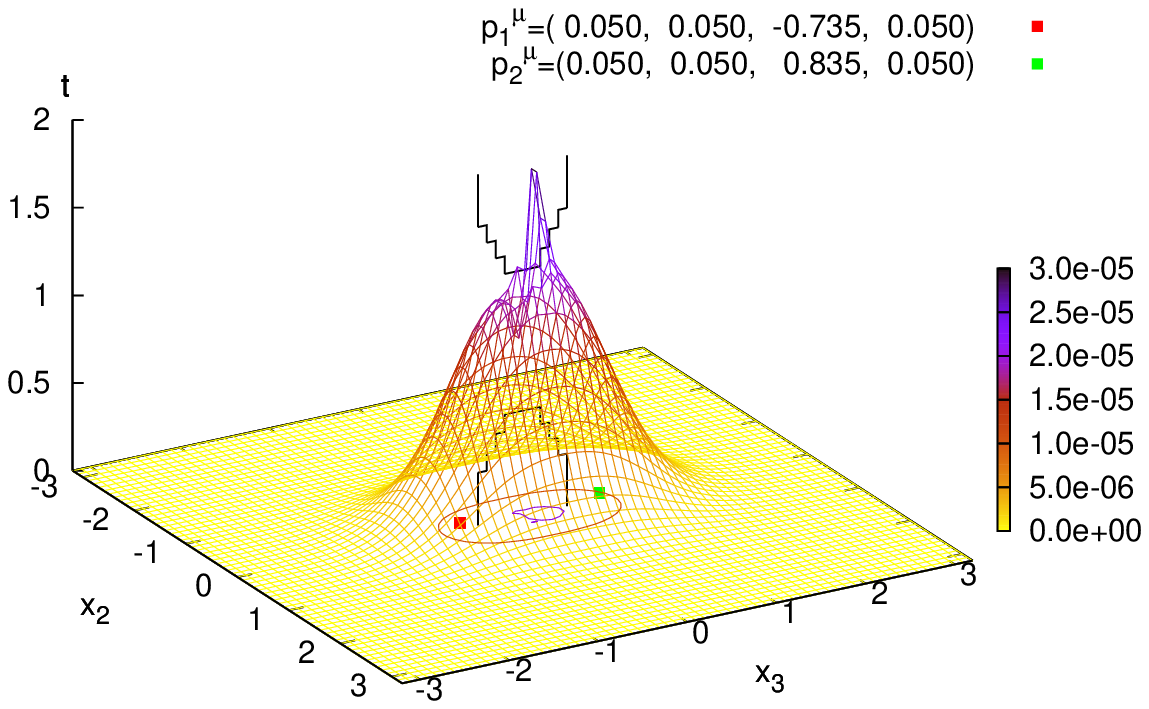}}%
  \put(4900,2500){(b)}
  \linethickness{0.09mm}
  \qbezier(5120,3980)(5270,3680)(5400,4030)
  \put(5400,4030){\vector(1,3){5}}
  \qbezier(5120,3550)(5270,3850)(5400,3600)
  \put(5120,3550){\vector(-1,-2){5}}
  \put(0,00){\includegraphics[trim=0 0 0 0, width=90mm]%
           {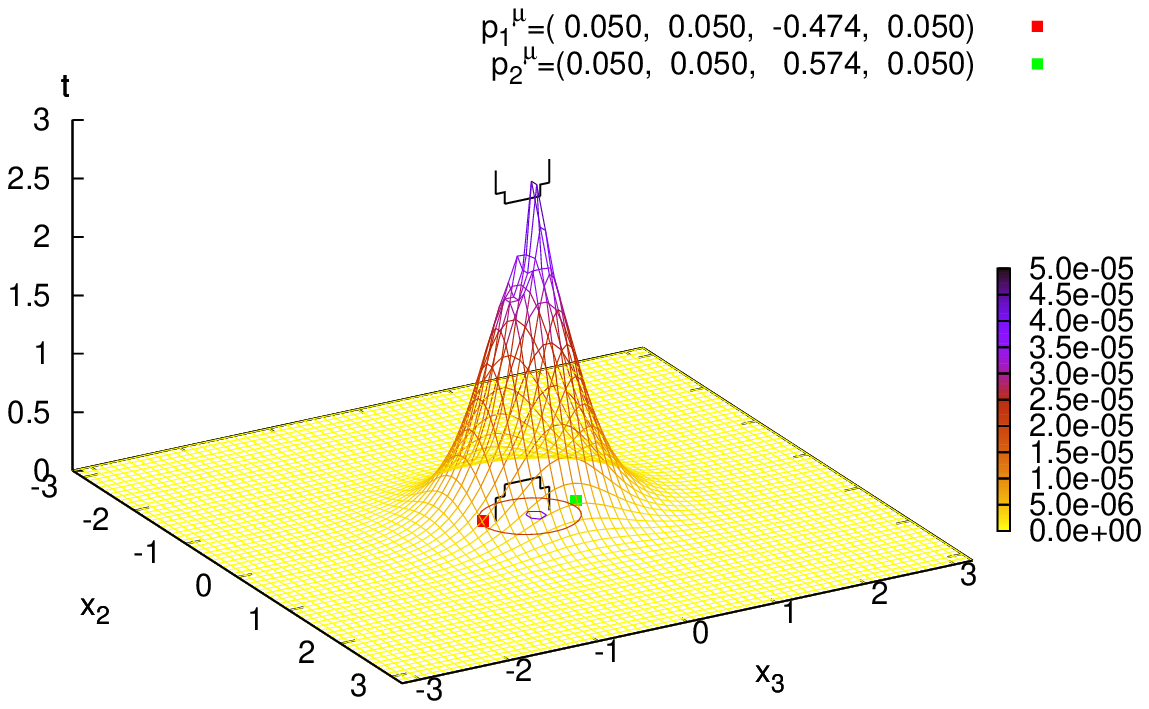}}%
  \put(1400,00){(c)}
  \put(3500,00){\includegraphics[trim=0 0 0 0, width=90mm]%
              {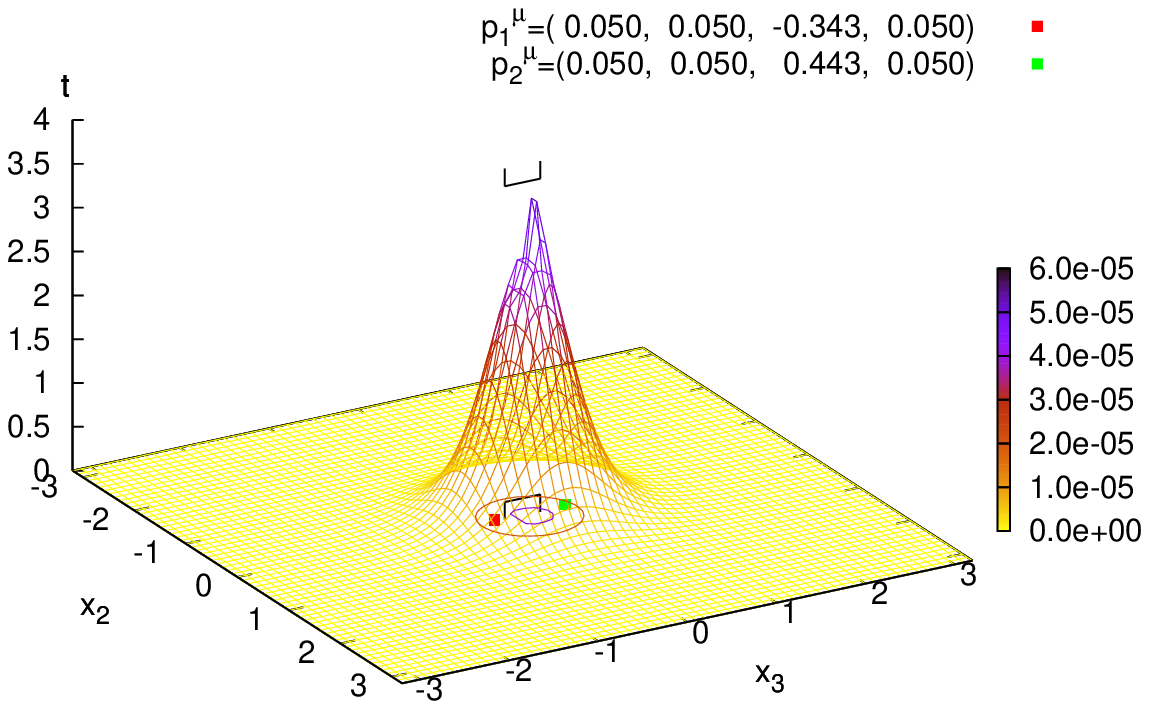}}%
  \put(4900,00){(d)}
 \end{picture}
 \caption{The charge distribution $D_x$ of the KvBLL caloron with a fixed $\rho=10a=1.0$  and
the associated magnetic--monopole current $k_{x,\mu}$  in the three-dimensional space $x_1=0$ projected from the four-dimensional space, which is regularized by the lattice of a finite volume $V$ with a fixed $L_x=35$ and various choices of $L_t$: (a)$L_t=10$, (b)$L_t=20$, (c)$L_t=30$,
(d)$L_t=40$.
The grid shows the caloron charge density $D_x$ on the $x_2$-$x_3$ ($x_1=t=0$) plane, together with its value being indicated in the legend on the right  side.
The black (thick) curve on the $x_2$-$x_3$ plane shows the  loop of the magnetic current $k_{x,\mu}$ and the arrow indicates the direction of the monopole current. 
The colored (thin) lines on the   $x_2$--$x_3$ plane show a contour plot for the equi-$D_x$ lines.
Due to the  periodicity of the $t$ direction with the period $\beta=aL_t$, the top plane $t=\beta$ should be identified with the bottom one $t=0$.
Here $p_1, p_2$ denote the location of the poles of the KvBLL caloron.
}
 \label{fig:nontrivial-caloron_loop}
\end{figure*}

First, we consider the HS caloron.
The result of numerical calculations for the HS caloron are summarized
in  FIG.~\ref{fig:trivial1carolon}  and TABLE~\ref{table:trivial1carolon}.

\begin{table}[htbp]
 \begin{center}
  \begin{tabular}{c||c|c||cccc||c}
   &$L_t$&$\beta$&$|k_{x,\mu}|>1$&$k_{x,\mu}=-1$&$k_{x,\mu}=0$&$k_{x,\mu}=1$&$Q_V$\\
   \hline
    (a)&10&1.0&0&4&13729992&4&0.952\\
   \hline
    (b)&20&2.0&0&8&13729984&8&0.975\\
   \hline
    (c)&30&3.0&0&8&13729984&8&0.980\\
   \hline
    (d)&40&4.0&0&8&13729984&8&0.983\\
  \end{tabular}
 \caption{The distribution of generated configurations of $k_{x,\mu}$ and the charge $Q_V$ for the HS caloron on the lattice with a volume $V=(2aL_x)^3 aL_t$ with fixed $L_x=35$ and various $L_t=10,20,30,40$ (with $a=0.1$).}
 \label{table:trivial1carolon}
\end{center}
\end{table}

TABLE~\ref{table:trivial1carolon} shows that one-caloron is approximately constructed on the lattice with a finite volume $V$, which is confirmed by 
\begin{equation}
 Q_V=0.952 \cong 1 ,
\end{equation}
 for the choice of the lattice:
\begin{equation}
L_x=35, \quad L_t=10,20,30,40 .
\end{equation}


For this choice, the total number of configurations for the magnetic current $k_{x,\mu}$
is $(2L_x)^3 \times L_t \times 4=70^3 \times 10 \times 4=13720000$.
TABLE~\ref{table:trivial1carolon} shows 
the distribution of the resulting magnetic current $k_{x,\mu}$ on the lattice, which means that the current $k_{x,\mu}$ is zero on almost all the links $(x,\mu)$ except for a small number of links, in fact, $k_{x,\mu}$ is non-zero only on  4+4 links for $L_t=10$ and 8+8 links for $L_t=20,30,40$ independently of $L_t$.

FIG. \ref{fig:trivial1carolon} shows that the links on which $k_{x,\mu}$ has non-zero value are localized at the center 
of the HS caloron. 
The HS caloron reduces to one-instanton of 't Hooft type in the limit $\beta \rightarrow \infty$. 
Therefore, this result is consistent with that for one-instanton of 't Hooft type obtained in our past study \cite{FKSS10}.
Thus, we conclude that the HS caloron with a trivial holonomy does not lead to a nontrivial loop of magnetic monopole.

\subsection{KvBLL caloron}

\begin{table}[htbp]
 \begin{center}
\begin{tabular}{c||c|c||cccc||c}
  &$L_t$&$\beta$&
  $|k_{x,\mu}|>1$&$k_{x,\mu}=-1$&$k_{x,\mu}=0$&$k_{x,\mu}=1$&
  $Q_V$\\
  \hline
   (a)&10&1.0& 0&10&13719980&10&0.973\\
  \hline
   (b)&20&2.0& 0&22&13719956&22&0.986\\
  \hline
   (c)&30&3.0& 0&12&13719976&12&0.987\\
  \hline
   (d)&40&4.0& 0&8&13719984&8&0.987
\end{tabular} 
 \end{center}
 \caption{The distribution of generated configurations of $k_{x,\mu}$ and the charge $Q_V$ for the KvBLL caloron on the lattice with  a volume $V=(2aL_x)^3 aL_t$ with fixed $L_x=35$ and various $L_t=10,20,30,40$ (with $a=0.1$).}
 \label{table:nontrivial-caloron_loop}
\end{table}

\begin{figure}[htbp]
 \begin{center}
  \includegraphics[trim=15 0 0 0, width=85mm]{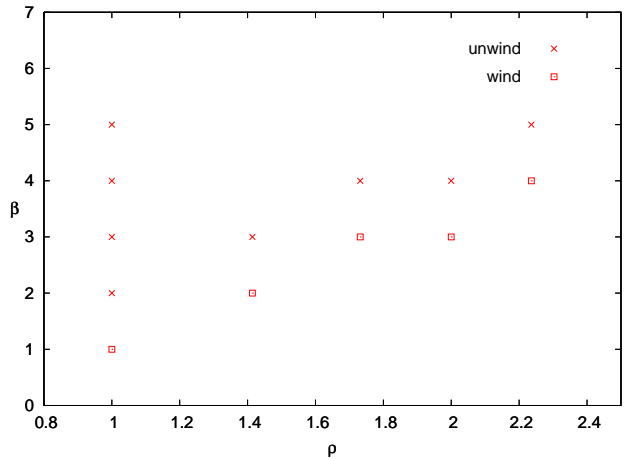}
 \end{center}
 \caption{The magnetic-monopole loops generated from
the KvBLL caloron wind or unwind around $S^1$ depending on the choices of $\rho$ and $\beta$. The winding case is denoted by a circle, while unwinding by the cross.
}
 \label{fig:positive_correlation}
\end{figure}

Next, we consider the KvBLL caloron.
The result of numerical calculations for the KvBLL  caloron are summarized
in FIG.~\ref{fig:nontrivial-caloron_loop} and TABLE~\ref{table:nontrivial-caloron_loop}.

For the KvBLL caloron,
the function $h(x)$ in Eq.\eqref{BehaviorOfA} is $h(x)=\bm{1}$, therefore,
a boundary condition for ${\bf n}_x$ is given by (\ref{n-boundary}).
We prepare the KvBLL caloron with the parameters $\rho$ and $x_0^\mu=(0,0,0,0)+\Delta$ on a lattice with a fixed lattice spacing $a=0.1$ with the size:
a fixed 
$L_x$  and various choices for $L_t$: 
\begin{equation}
L_x=35, \quad L_t=10,20,30,40 .
\end{equation}
The nontrivial holonomy $H$ is fixed by taking 
\begin{equation}
 \theta =\pi.
\end{equation}
Here, we have calculated the magnetic monopole by changing $L_t$,
since we are interested in behavior of the resulting magnetic monopole loop for variation of the period $\beta=aL_t=1.0, 2.0, 3.0, 4.0$.



First, we fix $\rho=1.0$.
TABLE~\ref{table:nontrivial-caloron_loop} and FIG.~\ref{fig:nontrivial-caloron_loop} 
show the resulting magnetic monopole loops generated from the KvBLL caloron.
FIG.\ref{fig:nontrivial-caloron_loop} shows how  the magnetic monopole loops behave as $\beta$ changes. 
In FIG.\ref{fig:nontrivial-caloron_loop}(a) for $\beta=1.0$, two magnetic loops wind along $S^1$ and each of the two loops passes through a pole of the KvBLL caloron.
Here the magnetic monopole current  $k_{x,\mu}$ runs in opposite directions.

As $\beta$ becomes larger, see   (b) for $\beta=2.0$, two poles $p_1, p_2$ get together and two loops fuse into a trivial loop (with a trivial winding number) which is not winding along $S^1$, before two currents in the opposite directions collide to be annihilated.
Eventually, the loop tends to shrink towards the center of the caloron, as seen in (c) for $\beta=3.0$ and (d) for $\beta=4.0$.

Our results indicate that the KvBLL caloron with nontrivial holonomy can be the source of the loop of magnetic monopole.

Next, we change $\rho$ in the range $\rho^2=1,2,3,4,5$.
FIG.~\ref{fig:positive_correlation} shows whether the resulting magnetic loop winds or not along $S^1$ for various choices  of $L_t$ and $\rho$.
FIG.~\ref{fig:positive_correlation}  indicates that the critical circumference $\beta_c$
at which the winding number of the loop varies exists and that $\beta_c$ depends on $\rho$: 
$\beta_c$ and $\rho$ have a positive correlation, which is schematically shown in FIG.~\ref{fig:positive_correlation}.
$\beta_c$ is approximately proportional to $\rho$: 
\begin{equation}
 \beta_c \propto \rho .
\end{equation}
It turns out that the loop winds along $S^1$ in smaller $\beta$ for smaller $\rho$ and that the critical is non-zero $\beta_c>0$ for any value of $\rho$. 
Such magnetic monopole loops as those shown in this paper had been shown to exist on the lattice using lattice simulations \cite{Ejiri96,CZ07,AE08}.
The $\beta$ dependence of  magnetic-monopole loops generated from the instanton with trivial holonomy was studied in a different formulation, see \cite{STSM96}.

\section{Summary}

In this paper, we have investigated the possible magnetic monopole content in the one-caloron solution, i.e., a periodic self-dual solution of the Yang-Mills field equation with the period $\beta$ defined on $\mathbb{R}^3\times S^1$.
Using the method developed in our previous paper \cite{FKSS10}, we have shown in a numerical way that the one-caloron solution with nontrivial holonomy, i.e., KvBLL caloron, can be a source of the closed loop of magnetic monopoles, while the one-caloron with trivial holonomy, i.e., HS caloron, does not generate the magnetic monopole loop.

The magnetic loop generated from the KvBLL caloron changes its topological behavior depending on the magnitude of the periodicity $\beta$, which is the length of the circumference of $S^1$ in $\mathbb{R}^3\times S^1$. 
Since the $\beta$ is identified with the inverse temperature $T^{-1}$ in the Yang-Mills theory at finite temperature, this result could be a clue to understand the phase transition from confinement phase to the deconfinement phase at finite temperature from the viewpoint of magnetic monopole according to the dual superconductor picture for the QCD vacuum, as studied based on dyons in \cite{DP}. 
The detailed study from this viewpoint will be given in future publications.

{\it Acknowledgements}\ ---
This work is  supported by Grant-in-Aid for Scientific Research (C) 21540256 from Japan Society for the Promotion of Science (JSPS), and also in part by the JSPS Grant-in-Aid for Scientific Research (S) \#22224003. 
The numerical calculations are supported by the Large Scale Simulation Program No.09-15 (FY2009) and No.10-13 (FY2010) of High Energy Accelerator Research Organization (KEK).


\end{document}